\documentclass[nonacm, sigplan, 9pt]{acmart}

\usepackage{amsmath,amssymb,amsfonts}
\usepackage{algorithmic}
\usepackage{graphicx}
\usepackage{textcomp}
\usepackage{xcolor,soul}
\usepackage{subfig}
\usepackage{todonotes}
\usepackage{mdframed}
\usepackage{hyperref}
\usepackage{multirow}
\usepackage{xspace}

\AtBeginDocument{%
  }

\renewcommand\footnotetextcopyrightpermission[1]{}
\newcommand{\lumina}{\textsc{Lumina}\xspace}
\newcommand{\eg}{\emph{e.g.,}\xspace}
\newcommand{\equalcontribfootnote}{%
  \begingroup
  \renewcommand{\thefootnote}{\fnsymbol{footnote}}%
  \setcounter{footnote}{0}%
  \footnotetext[1]{Both authors contributed equally to this research.}%
  \endgroup
}

\begin{document}

\title{LUMINA: LLM-Guided GPU Architecture Exploration via Bottleneck Analysis}

\author{Tao Zhang$^{*}$, Rui Ma$^{*}$, Shuotao Xu, Yongqiang Xiong, Peng Cheng}
\affiliation{%
  \institution{Microsoft Research}
  \country{China}
}
\email{{zhangt, mrui, shuotaoxu, yqx, pengc}@microsoft.com}

\begin{abstract}
GPU design space exploration (DSE) for modern AI workloads, such as Large-Language Model (LLM) inference, is challenging because of GPUs' vast, multi-modal design spaces, high simulation costs, and complex design optimization objectives (\eg performance, power and area trade-offs).
Existing automated DSE methods are often prohibitively expensive, either requiring an excessive number of exploration samples or depending on intricate, manually crafted analyses of interdependent critical paths guided by human heuristics.

We present \lumina, an LLM-driven GPU architecture
exploration framework that leverage AI to enhance the DSE efficiency and efficacy for GPUs. 
\lumina extracts architectural knowledge from
\emph{simulator code} and performs \emph{sensitivity studies} to automatically compose DSE rules,which are auto-corrected during exploration. A core component of \lumina is a DSE Benchmark that comprehensively evaluates and enhances LLMs’ capabilities across three fundamental skills required for architecture optimization,
which provides a principled and reproducible basis for model selection and ensuring \emph{consistent architectural reasoning}.

In the design space with 4.7 million possible samples, \lumina identifies 6 designs of better performance and area than an A100 GPU efficiently, using only 20 steps via LLM-assisted bottleneck analysis.
In comparison, \lumina achieves 17.5$\times$ higher than design space exploration efficiency, and 32.9\% better designs (i.e. Pareto Hypervolume) than Machine-Learning baselines, showcasing its ability to deliver high-quality design guidance with minimal search cost.

\end{abstract}

\maketitle
\equalcontribfootnote

\section{Introduction}

Graphics Processing Units (GPUs) have become the cornerstone of modern computing infrastructure in the age of artificial intelligence (AI). 
With global data center spending projected to exceed \$1 trillion by 2029~\cite{delloro2025datacenter}, largely driven by AI compute, optimizing GPU architecture for major workloads—especially AI training and inference—is critical to reducing total cost of AI infrastructure ownership and improving sustainability.

\begin{figure}[!t]
    \centering
    \subfloat[TTFT (s)]{%
    \includegraphics[width=0.3\linewidth]{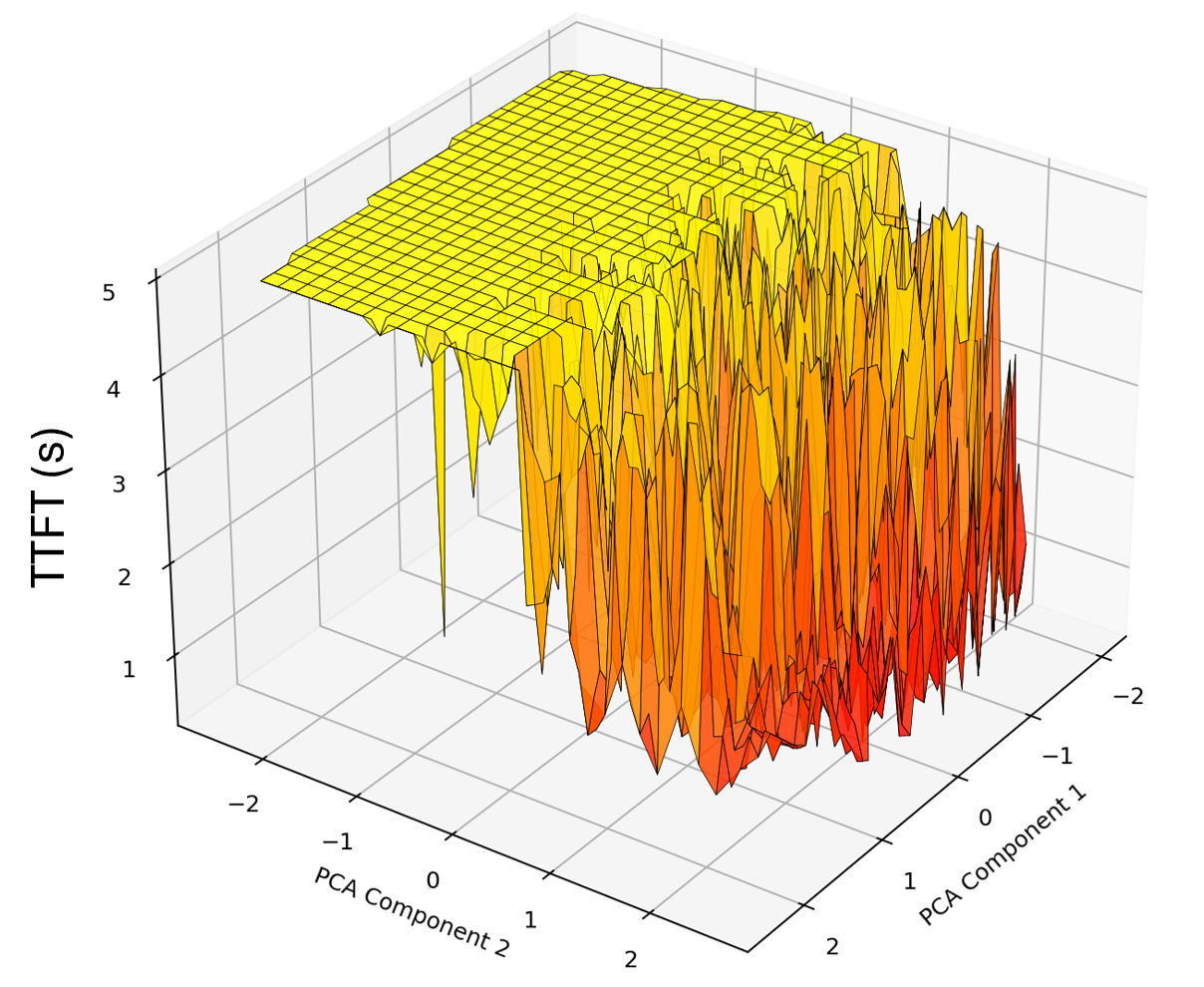}
    \label{fig:1a}
    }\hfill
    \subfloat[TPOT (s)]{%
    \includegraphics[width=0.3\linewidth]{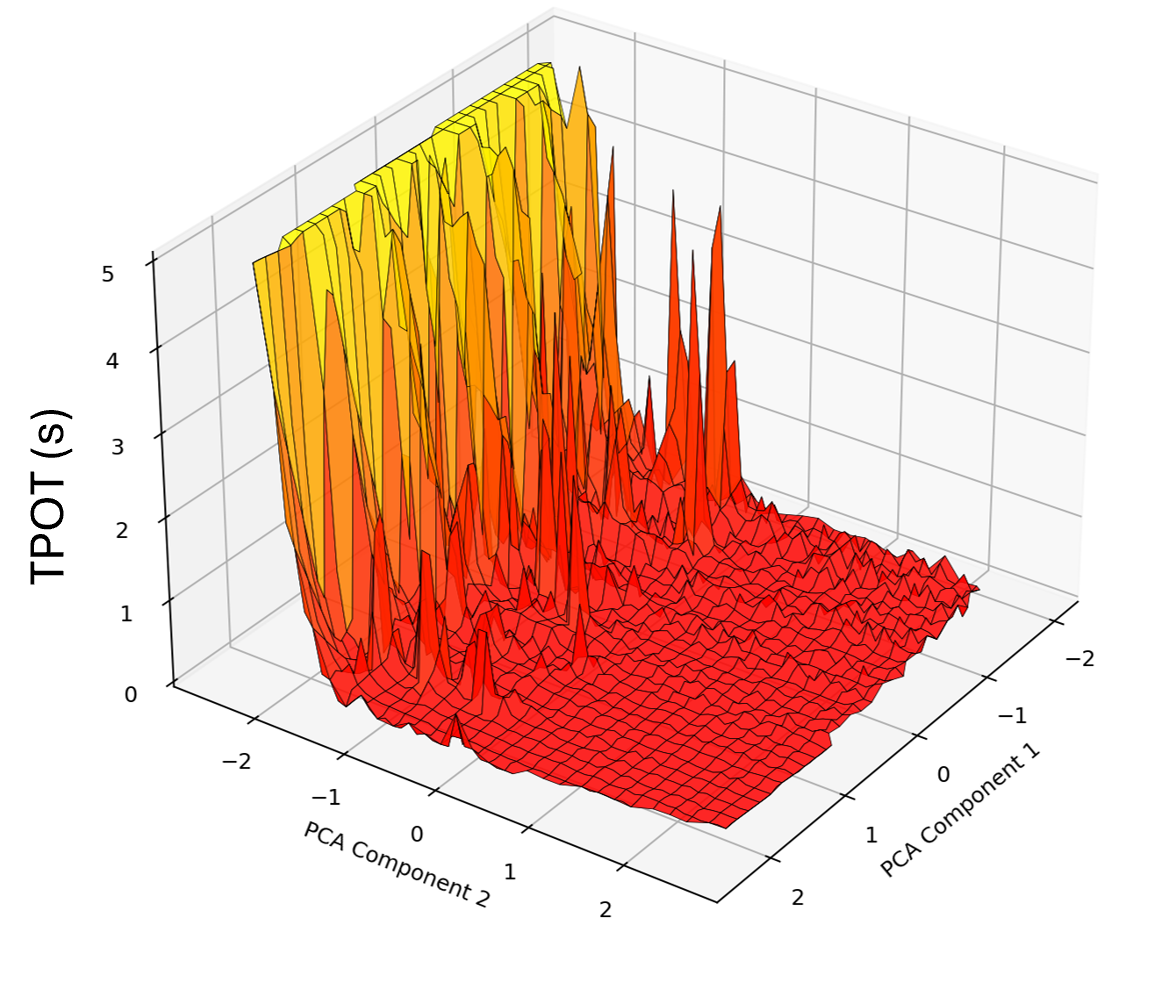}
    \label{fig:1b}
    }\hfill
    \subfloat[Area ($\mathrm{mm}^2$)]{%
    \includegraphics[width=0.3\linewidth]{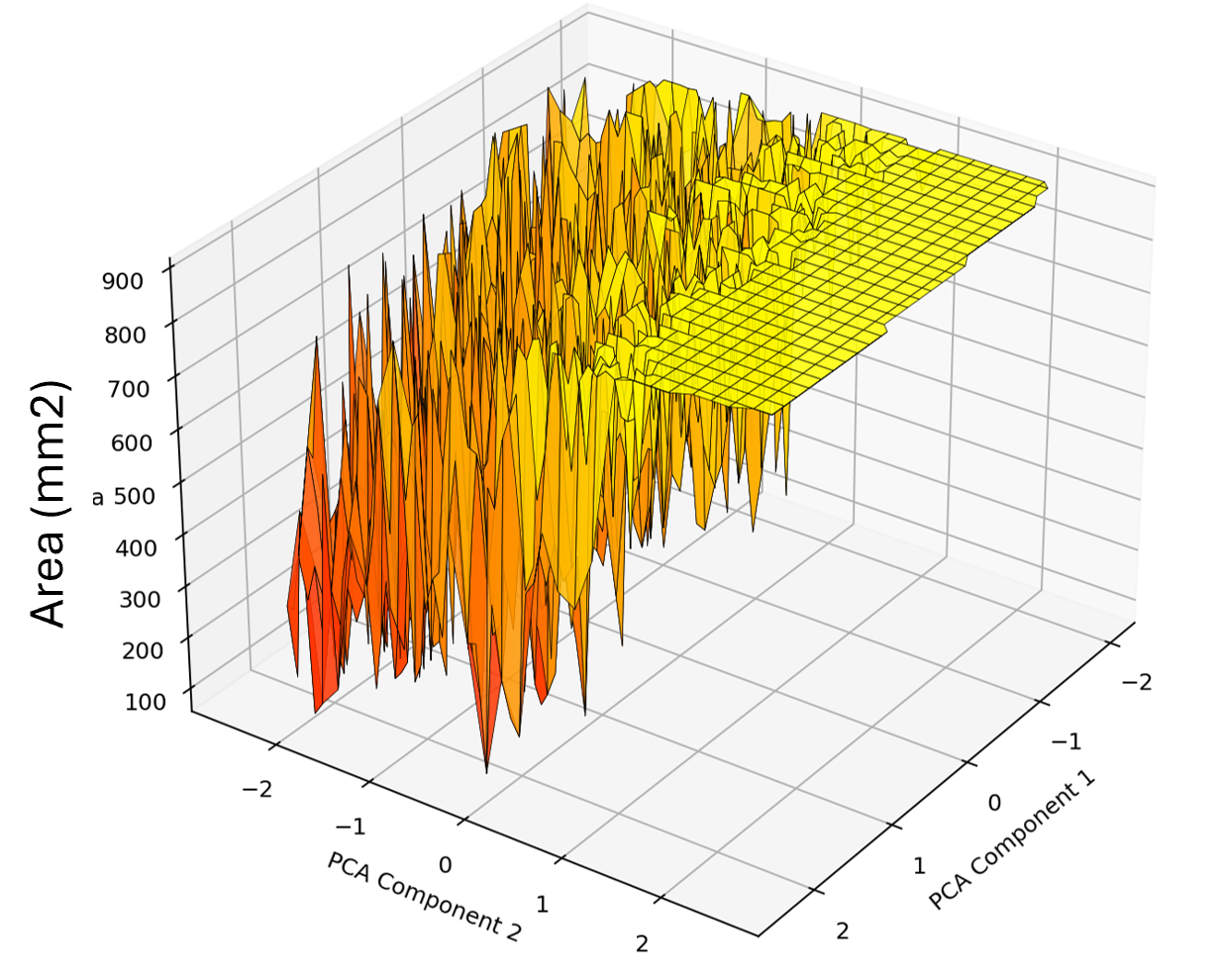}
    \label{fig:1c}
    }\hfill

    \caption{Design space visulizatin for a GPT3-175B~\cite{gpt3} inference workload via roofline model~\cite{roofline}. Each sampled architecture is embedded via Principal Component Analysis~\cite{pca} into two dimensions; distributions are capped for visual contrast.}
    \label{fig:design_space}
\end{figure}

However, optimizing GPUs via design space exploration (DSE) is often complex and time-consuming.
For example, exploring 1000 GPU designs for one GPT-3 inference workload trace takes 6000 CPU hours via state-of-the-art GPU analytical modelling~\cite{llmcompass2024isca}.
GPU DSE is difficult for three reasons.
First, the design space is intrinsically high-dimensional—spanning compute units, cache hierarchy, interconnects, and memory bandwidth—resulting in multi-modal parameter distributions that hinder efficient search.
Second, evaluation is costly, as each candidate typically requires hours of detailed simulation.
Finally, the objective landscape is also multi-modal: performance, power, and area interact non-linearly, producing complex Pareto fronts that complicate trade-off decisions.
Figure~\ref{fig:design_space} illustrates these inherent challenges of the GPU DSE problem for LLM inference workloads. The design space comprises approximately 4.7 million candidate architectures, each associated with distinct design objectives—including Time-to-First-Token (TTFT), Time-Per-Output-Token (TPOT)~\cite{mlcommons2025llmInferenceV5}, and chip area. These objectives frequently conflict with one another, resulting in intricate trade-offs that further complicate the identification of an optimal design.

Existing DSE methods often suffer from low exploration efficiency, typically requiring hundreds to thousands of samples to find high-quality designs. They fall into two broad categories: expert-driven heuristics~\cite{exp2015hpca,exp2012micro,exp2011micro} and algorithmic ML-based tools~\cite{tool2023dac,tool2024vlsi,tool2024dac}.
Expert-based approaches use rules such as critical-path analysis and stall-component mapping to identify bottlenecks, achieving reasonable Pareto hyper-volume (PHV) with a few hundred simulations. 
However, their heuristic nature requires substantial human domain expertise, limits generalization to new architectures, and makes it difficult to capture complex interactions across multiple critical paths~\cite{cost2003micro}.
ML-based methods aim to address these limitations by learning the non-linear structure of the design space and can reveal insights beyond static heuristics. Yet, they require large numbers of high-fidelity simulation samples for training, making them expensive to deploy.

Recent explosion in AGI capabilities offers a silver lining for advancing DSE. Emerging studies show that LLMs can enhance ML-based DSE by leveraging their pretrained domain knowledge to guide sample selection~\cite{evo2023nips,ramos2025bayesianoptimizationcatalysisincontext,lemoe25dac}. This potential builds on LLMs’ demonstrated strengths in coding~\cite{evo2023nips}, reasoning~\cite{wang2024q}, and applying domain-specific knowledge~\cite{liu2025datasets}. However, LLM performance varies widely across models~\cite{lmarena2025leaderboard}, and issues such as hallucination~\cite{tonmoy2024comprehensivesurveyhallucinationmitigation} and incomplete domain understanding~\cite{liu2025datasets} limit their reliability. These challenges underscore the need for a systematic and rigorous methodology to assess LLMs’ effectiveness for DSE, rather than relying on ad-hoc or anecdotal evaluations.

In this paper we propose \lumina, an LLM-driven GPU architecture exploration framework that delivers \emph{reliable} and \emph{sample-efficient} architectural reasoning, addressing both the generality limits of human-crafted heuristics and the instability of vanilla LLM agents. \lumina acquires architectural knowledge directly from simulator code and sensitivity studies, and then uses this knowledge to construct exploration rules that are subsequently auto-corrected as new samples are observed.

At the core of \lumina is a DSE Benchmark that evaluates three fundamental LLM capabilities essential for architecture optimization: bottleneck attribution, performance/area prediction, and parameter tuning. This benchmark systematically exposes LLM weaknesses in DSE and provides a \emph{principled, reproducible} basis for selecting LLMs capable of consistent architectural reasoning—moving beyond ad-hoc, model-dependent usage.

We make the following contributions in this paper:

\begin{itemize}

\item \textbf{\lumina Framework}:
An LLM-guided GPU DSE framework that is both reliable and sample-efficient. Using NVIDIA A100 as reference, \lumina achieves 32.9\% higher PHV than ML baselines, improves sample efficiency by up to 17.5×, and is the only method to find superior designs within 20 samples across a 4.7M-point search space.

\item \textbf{DSE Benchmark}:
The first benchmark for LLM-driven GPU DSE, assessing three essential capabilities for chip architectural reasoning: bottleneck attribution, performance/area prediction, and parameter tuning. It exposes systematic failure cases in LLM-based analysis and offers a reproducible mechanism for selecting LLMs that reason consistently within \lumina.

\item \textbf{New DSE Strategy}:
\lumina uncovers a counter-intuitive DSE strategy: reallocating area from core counts to tensor-compute units and memory bandwidth improves overall PPA. Via this strategy, \lumina uncovers two superior designs than A100: one achieves 1.805$\times$ TTFT/Area and 1.770$\times$ TPOT/Area, while another prioritizes TTFT with 0.592$\times$ TTFT and 0.948$\times$ TPOT than the baseline, both with reduced area.

\end{itemize}

\noindent\textbf{Paper Organization:} Section \ref{sec:background} reviews existing DSE methods. Section \ref{sec:framework} presents the \lumina framework, and Section \ref{sec:bottleneck_reasoning_benchmark} describes the DSE Benchmark. Section \ref{sec:eval} details our implementation and evaluation results, and Section \ref{sec:conclusion} concludes the paper.

\section{Background}
\label{sec:background}

\subsection{DSE Problem Formulation}

A typical DSE task over a multi-objective design space can be formulated as follows:

\textbf{Definition 1} (Design Space):
The design space of a GPU node is detailed in Table~\ref{tab:gpu_design_space} and denoted by $\mathcal{X}$, where each design point is represented as $\mathbf{x} \in \mathcal{X}$.

\begin{table}[htbp]
\centering
\setlength{\tabcolsep}{5pt}
\caption{Example Design Space of a 8 GPU Node}
\label{tab:gpu_design_space}
\begin{tabular}{|l|p{3cm}|c|}
\hline
\textbf{Parameter} & \textbf{Value Range} & \textbf{\#} \\
\hline
Interconnect Link Count & 6, 12, 18, 24 & 4 \\
\hline
Core Count & 1, 2, 4, 8, 16, 32, 64, 96, 108, 128, 132, 136, 140, 256 & 14 \\
\hline
Sublane Count & 1, 2, 4, 8 & 4 \\
\hline
Systolic Array Height and Width & 4, 8, 16, 32, 64, 128 & 6 \\
\hline
Vector Width & 4, 8, 16, 32, 64, 128 & 6 \\
\hline
SRAM Size (KB) & 32, 64, 128, 192, 256, 512, 1024 & 7 \\
\hline
Total Global Buffer (MB) & 32, 64, 128, 256, 320, 512, 1024 & 7 \\
\hline
Memory Channel Count & 1--12 & 12 \\
\hline
\textbf{Total Design Points} & \multicolumn{2}{c|}{Approximately $4.7 \times 10^{6}$} \\
\hline
\end{tabular}
\end{table}

\textbf{Definition 2} (Pareto Optimality):
A design point $\mathbf{x}^*$ is \textit{Pareto-optimal} if 
I
improving any objective requires sacrificing at least one other objective. The set of all Pareto-optimal points forms the \textit{Pareto frontier}.

\textbf{Definition 3} (Pareto Hypervolume, PHV):
The Pareto Hypervolume measures the quality of a \textit{Pareto frontier} by computing
the $m$-dimensional volume dominated by a set of Pareto-optimal points with
respect to a given reference point. A larger PHV indicates a better
Pareto frontier.

\textbf{Problem Formulation}: Given a GPU node design space $\mathcal{X}$, the goal of DSE is to identify the pareto-optimal configurations $\mathbf{x} \in \mathcal{X}$ upon multiple objectives, such as performance, power, and area and maximize PHV in a given time, measured by sample numbers. 

\begin{table*}[!t]
\centering
\caption{Comparison of representative DSE methods.}
\label{tab:dse_methods}
\setlength{\tabcolsep}{5pt}
\renewcommand{\arraystretch}{1}
\begin{tabular}{|c|l|l|l|l|}
\hline
\textbf{Category} & \textbf{Representative Method} & \textbf{Sample Learning} & \textbf{Sample Efficiency} & \textbf{Sample Scalibity} \\
\hline
\multirow{2}{*}{Heuristic} & Grid Search~\cite{gs2012} & No & Random & High \\
\cline{2-5}
 & Random Walker~\cite{rw2020} & No & Random & High \\
\hline
\multirow{4}{*}{Machine Learning} & Bayesian Optimization (BO)~\cite{lemoe25dac} & Yes & Low~\cite{archgym23isca} & Low\cite{bayes2021nips} \\
\cline{2-5}
 & Genetic Algorithms (GA)~\cite{gamma202iccad} & Yes & Low~\cite{archgym23isca} & Low\cite{ga2015scale} \\
\cline{2-5}
 & Ant Colony Optimization (ACO)~\cite{aco2021iscas} & Yes & Low~\cite{archgym23isca} & Low\cite{aco2019scale} \\
\hline
\multirow{2}{*}{Expertise-Driven} & Bottleneck-Removal via Critical Path Analysis~\cite{archexplorer2023micro} & No & High & Low \\
\cline{2-5} 
& \textbf{\lumina (Our work)} & \textbf{Yes} & \textbf{High} & \textbf{Medium} \\
\hline
\end{tabular}
\end{table*}

\subsection{Existing DSE Methods}

Existing DSE methods can be broadly categorized into \textit{black-box} approaches (including heuristic and machine-learning-based methods) and \textit{white-box} approaches (driven by expert knowledge), as summarized in Table~\ref{tab:dse_methods}. 

Heuristic methods such as Grid Search~\cite{gs2012} and Random Walker~\cite{rw2020} do not exploit prior samples or exploration knowledge, leading to uncontrolled variance and random sampling efficiency in design quality~\cite{archgym23isca}. 
In contrast, machine learning-based approaches such as Bayesian Optimization (BO), 
Genetic Algorithms (GA), and Ant Colony Optimization (ACO) aim to identify promising design points through model-based or policy-driven feedback~\cite{archgym23isca, boom2021ICCAD, lemoe25dac}, a process we refer to as \textit{sample learning}.
However, these methods generally suffer from \textit{low sample efficiency} and \textit{limited scalability} in high-dimensional DSE tasks~\cite{bayes2021nips, lemoe25dac, NEURIPS2018_f02208a0, jomaa2019hyprlhyperparameteroptimization}. The low sample efficiency arises from the sparsity of high-quality design points in the search space, necessitating more effective strategies to provide initial samples.
The scalability issue, on the other hand, stems from the computational and memory costs that can grow cubically with the number of samples~\cite{ga2015scale, aco2019scale}.

Specifically, BO relies on surrogate models to understand the PPA-resource mapping and utilizes acquisition function to improve toward optimization goal. However, its cubic time complexity leads to poor scalability with large design spaces~\cite{bayes2021nips}. 


GA performs evolutionary search through mutation and crossover but converges slowly, requiring over 10k samples in DNN DSE evaluations~\cite{gamma202iccad}. 
ACO explores via pheromone-guided probabilistic sampling~\cite{aco2021iscas}, exhibiting lower sample efficiency than BO and GA under various benchmarks~\cite{archgym23isca}. 
While distributed or adaptive variants have been proposed to improve scalability, large-scale evaluations still report limited improvement~\cite{ga2015scale, aco2019scale}.

Beyond black-box search, \textit{Critical Path Analysis}\cite{cpa2014micro} offers a white-box alternative that leverages expert knowledge of architectural bottlenecks. 
As a representative, Archexplorer~\cite{archexplorer2023micro} performs bottleneck-removal-driven DSE by reallocating resources along critical paths, achieving superior sample efficiency and the highest PHV under equal sampling budgets compared with black-box approaches. 
However, the bottleneck-to-resource mapping in such white-box approaches is heuristically defined, which prevents refinement through sampling and limits generality across architectures~\cite{rl2024aaai}.

LLMs offer a mechanism to achieve the strengths of both methods, the \emph{sample learning} capability and the high \emph{sample efficiency}, by reflecting on exploration trajectories and conducting human-like architectural reasoning to explore.
However, their capabilities have not been systematically evaluated
on DSE tasks. To address this gap, we design the \lumina framework
with a comprehensive DSE Benchmark, forming a principled framework
to assess model performance and enforce consistent architectural reasoning.
\section{\lumina Framework}
\label{sec:framework}
\newcounter{promptctr}
\renewcommand{\thepromptctr}{\arabic{promptctr}}

\newenvironment{promptbox}[1]{%
  \refstepcounter{promptctr}%
  \begin{mdframed}[%
      linecolor=black,
      linewidth=0.5pt,
      leftmargin=0pt,
      rightmargin=0pt,
      innertopmargin=2pt,
      innerbottommargin=2pt,
  ]%
  \textbf{Prompt-Answer \thepromptctr: #1}%
  \par
}{%
  \end{mdframed}%
}
\subsection{Overview of \lumina}

\begin{figure}[htbp]
    \centering
    \includegraphics[width=1\linewidth]{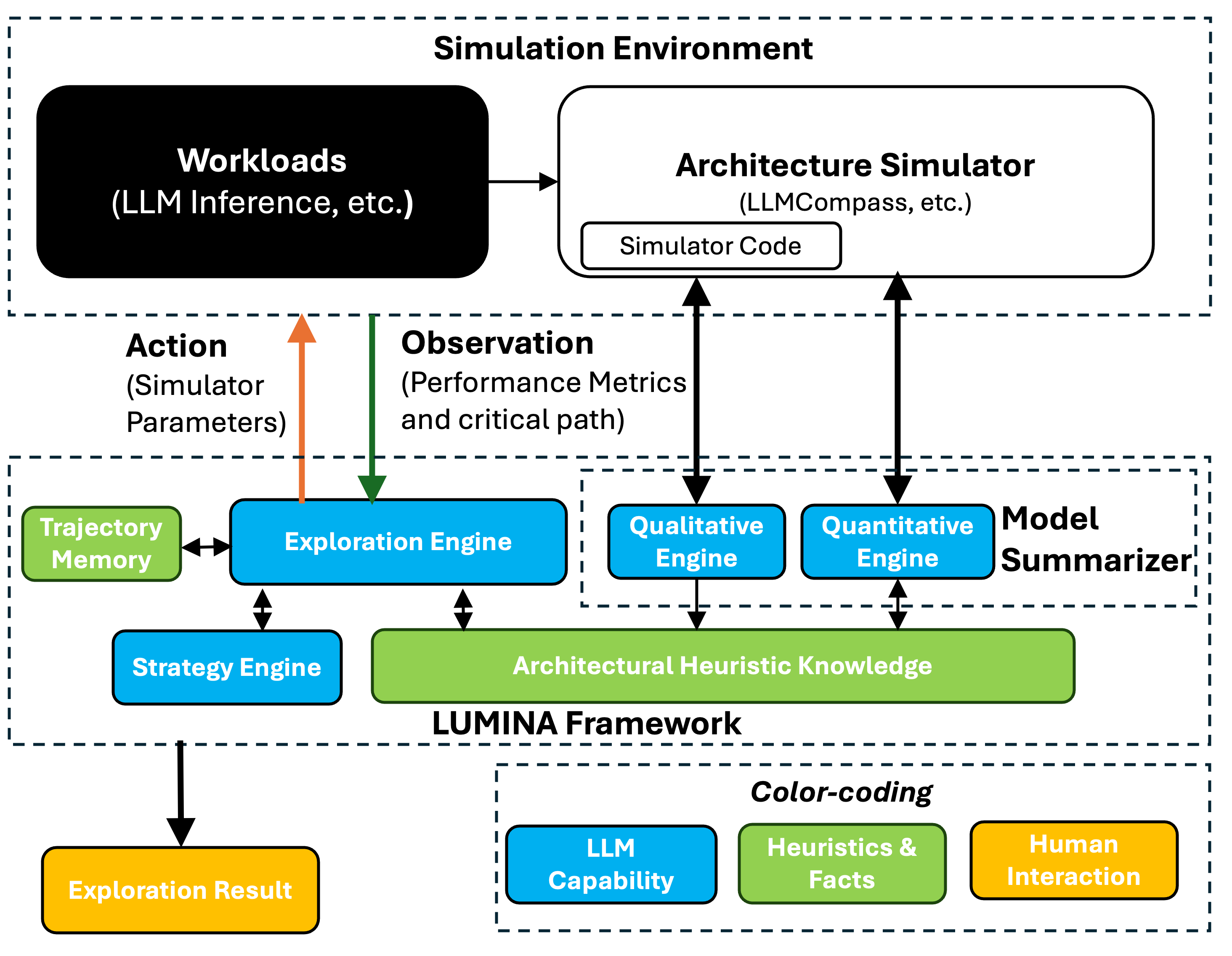}
    \label{fig:2}
    \caption{\lumina Framework Design}
    \label{fig:framework}
\end{figure}


\noindent Figure~\ref{fig:framework} presents the overall design of the \textbf{\lumina Framework}, which is structured around an iterative knowledge acquisition and refinement loop. \lumina begins by interacting with the external simulation environment to extract preliminary \textit{Architectural Heuristic Knowledge (AHK)}. This process is handled by the \textit{Qualitative Engine (QualE)}, which parses the simulator codebase to attribute resources to metrics, and the \textit{Quantitative Engine (QuanE)}, which quantifies each resource's impact on PPA.

The core of \lumina's optimization lies in its iterative process. Starting from an initial design, the framework evaluates the design and sends the results (e.g., PPA and critical-path data) to the \textit{Strategy Engine (SE)}. Utilizing the knowledge from AHK, SE performs bottleneck analysis on the results and generates a mitigation strategy for the \textit{Exploration Engine (EE)}, which proposes a new, informed design point. Meanwhile, the simulation results are stored in the \textit{Trajectory Memory (TM)}, which is used to \textbf{refine AHK} in subsequent iterations. This process repeats until the sampling budget is met, ultimately outputting a set of Pareto-optimal designs.

The subsequent subsections detail \lumina's core components: \S\ref{sec:ahk_acquisition} focuses on the \textbf{automatic acquisition of AHK} through the Qualitative and Quantitative Engines; \S\ref{sec:strategy_exploration} describes the utilization of AHK to guide DSE via the Strategy and Exploration Engines; and \S\ref{sec:refinement_loop} elucidates the \textbf{refinement loop} that enables continuous learning and cross-architecture scalability.

\subsection{Architectural Heuristic Knowledge (AHK) Acquisition}
\label{sec:ahk_acquisition}

The initial step in the \lumina framework is the automated acquisition of \textbf{AHK}, a critical capability that distinguishes our approach from conventional black-box and white-box methods. AHK serves as a structural and quantitative understanding of how architectural resources are attributed to PPA metrics. The acquisition is handled by two complementary modules: the $\mathrm{QualE}$ and the  $\mathrm{QuanE}$.

\subsubsection{Qualitative Engine ($\mathrm{QualE}$)}
\label{sec:quale}

The primary role of the $\mathrm{QualE}$ is to establish the \textbf{structural boundaries} of the AHK. It achieves this by leveraging the \textbf{semantic understanding capabilities of a LLM} to parse and interpret the simulator's complex codebase. Specifically, the $\mathrm{QualE}$ performs static code analysis, utilizing the LLM's interpretative strength to explicitly map the causal influence of each resource hyper-parameters onto specific PPA metrics. This process generates an Influence Map, which structurally defines the exact dependencies between a performance specification and the concrete architectural parameters within the design space. For instance, the map identifies that \textit{peak vector compute throughput} is influenced by core count, sublane count, and vector unit, but has no direct structural dependency on the tensor unit. This derived, structurally constrained knowledge drastically reduces the search space for subsequent quantitative analysis, providing a level of informed pruning analogous to the initial heuristics employed in traditional white-box methods.

\subsubsection{Quantitative Engine ($\mathrm{QuanE}$)}
\label{sec:quane}

Building upon the structural dependencies defined by the $\mathrm{QualE}$'s Influence Map, the $\mathrm{QuanE}$ is responsible for assigning \textbf{quantitative influence values} to the AHK relationships. The $\mathrm{QuanE}$ achieves this by executing an automated preliminary \textbf{sensitivity analysis} against the simulator, leveraging the LLM's \textbf{code generation and orchestration capability} to script and manage the necessary micro-benchmarks. By systematically observing the impact of granular changes (e.g., a $\pm 1$ unit perturbation in the Core Count) on the attributed PPA metrics (e.g., Area), the $\mathrm{QuanE}$ quantifies the local influence of each resource. This comprehensive quantification enables \lumina to initialize its exploration with robust, informed priors, providing a significant and measurable advantage over conventional black-box methods that must learn the entire PPA-resource mapping from scratch. Under complex performance models where measuring performance perturbations is costly, the $\mathrm{QuanE}$ can focus on estimating only power and area, which are faster to evaluate, while still providing informative priors for exploration.

\subsection{Strategy and Exploration Engine}
\label{sec:strategy_exploration}

The Strategy and Exploration Engines are responsible for leveraging the acquired AHK to guide the DSE toward Pareto-optimal fronts efficiently.

\subsubsection{Strategy Engine ($\mathrm{SE}$)}
The $\mathrm{SE}$ defines the \textbf{bottleneck mitigation strategy} based on critical-path feedback provided by the simulator. Upon identifying the most dominant performance stall (e.g., interconnect congestion or memory latency), the $\mathrm{SE}$ uses the AHK to propose a constrained set of design parameter adjustments aimed at alleviating that specific bottleneck. Crucially, the $\mathrm{SE}$ determines the \textbf{aggressiveness} of the search by deciding how many design parameters to modify simultaneously in the next iteration. For instance, if interconnect is the bottleneck, the $\mathrm{SE}$ suggests maximizing interconnect link count while simultaneously seeking a resource trade-off (e.g., reducing Core Count) based on the quantitative influence factors. This guided approach ensures that exploration is always purposeful and locally optimal.

\subsubsection{Exploration Engine ($\mathrm{EE}$)}

The $\mathrm{EE}$ serves as the integration layer between the Simulation Environment and \lumina’s internal modules. It serializes the $\mathrm{SE}$’s design directives into the simulator’s required format, issues the evaluation request, and retrieves the resulting performance and critical-path metrics. The $\mathrm{EE}$ then records this feedback in the $\mathrm{TM}$ and returns the structured sample to the $\mathrm{SE}$, enabling informed selection of subsequent design points.

\subsection{Refinement Loop}
\label{sec:refinement_loop}




The \textbf{Refinement Loop} enables \lumina to iteratively refine its heuristics and overcome the static limitations of conventional white-box methods.

In each iteration, it \textbf{reflects on the trajectory history} stored in the $\mathrm{TM}$ to identify past design attempts that failed to meet PPA targets and conclude the patterns to prevent their repetition.

This reflection drives the core refinement: observed performance data are used to update the \textbf{quantitative influence factors} in the AHK, calibrating the model's understanding of the PPA-resource mapping.  
Through these data-driven corrections, \lumina dynamically adapts to non-linear and complex behaviors in the design space, maintaining high \emph{sample efficiency} and scalability across diverse architectural configurations.

\section{DSE Benchmark}
\label{sec:bottleneck_reasoning_benchmark}
Being the backbone of \lumina framework, the LLM reasoning model needs to be carefully chosen or finetuned to ensure its ability to analyze the performance metrics and tune the parameters intelligently. Additionally, the model needs to properly follow the instructions from system prompts to (i) respect the design constraints and (ii) avoid circular reasoning. Therefore, we design a Q\&A based benchmark to evaluate a model's ability for architecture exploration purpose. 

\begin{figure}[htbp]
    \centering
    \includegraphics[width=1\linewidth]{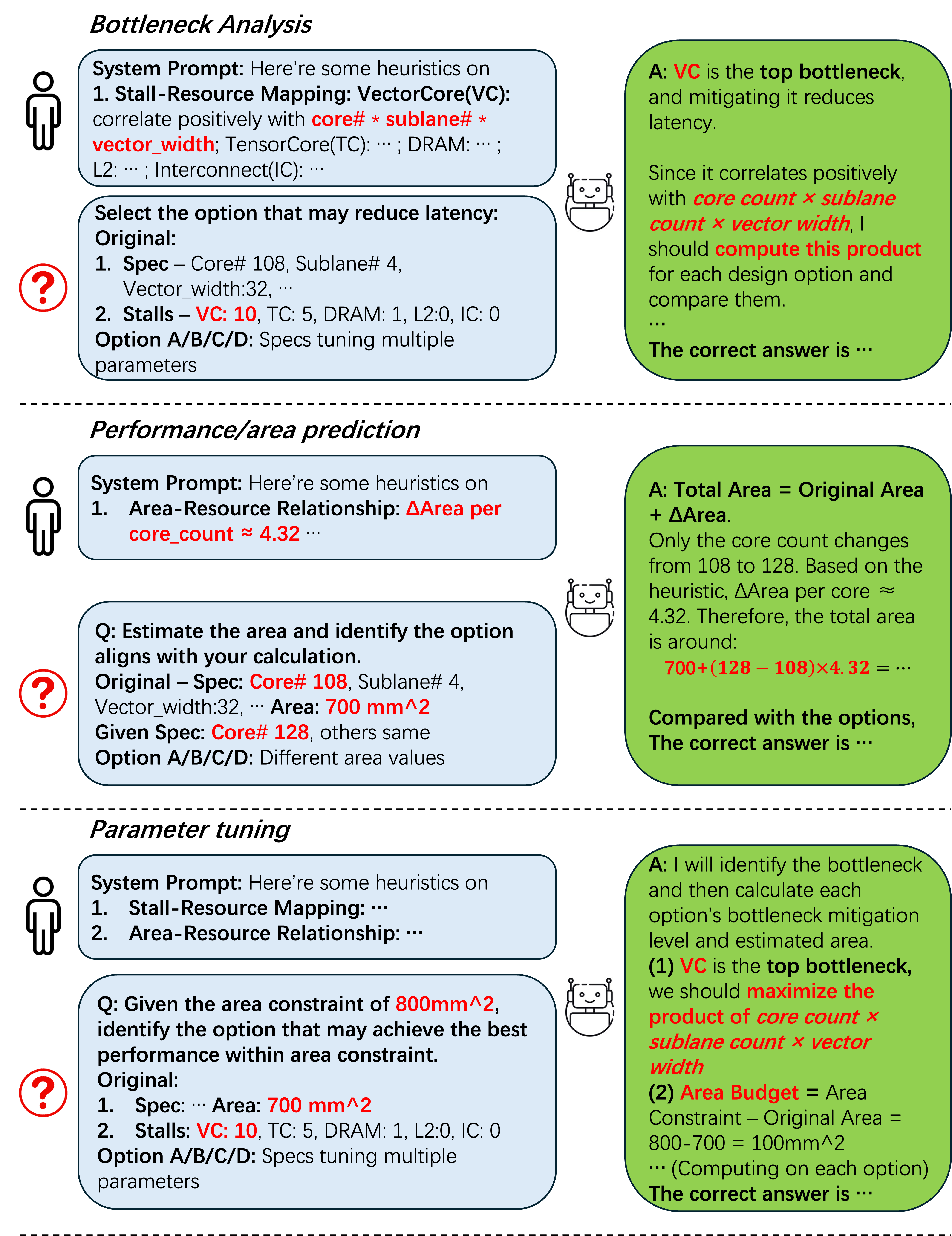}
    \caption{Examples of three task in DSE benchmark. For each task, we show the associated evidences from human system prompt, question on the left and LLM's correct answer on the right.}
    \label{fig:benchmark}
\end{figure}

Inspired by Longbench~\cite{bai2025longbenchv2}, we formulate each data sample as a multiple choice question. Each question starts with a system prompt and a description of the application target, ranging from primitive operators (e.g., matmul, layernorm, etc.) to full workload. A problem is then presented along with multiple-choice answers, of which only one is correct. The data samples can be categorized into three major tasks. Figure ~\ref{fig:benchmark} shows example data for each task.
\paragraph{Bottleneck analysis}
These questions evaluates the model’s fundamental capability to infer the relationship between performance counters and their corresponding architectural parameters. Each question prompt specifies the architectural configuration, the optimization objective, and the observed performance counter values for the target application. The model is required to determine which parameters should be adjusted and in which direction to achieve improvement toward the stated optimization goal.
\paragraph{Performance/area prediction}
These questions aim to assess the model’s capability to predict performance and area based on historical design trajectories and analytical models derived from source code. Each prompt provides several example architecture specifications along with their corresponding performance and area results, as well as the source code of the area model. The model is required to infer and select the correct performance or area value for a new architecture specification.
\paragraph{Parameter tuning}
These questions are intended to evaluate the model’s comprehensive capability in GPU architecture design space exploration. In addition to historical design trajectories and source code, each question specifies an initial design point, design constraints, and an optimization objective. The model is required to identify and select the design that best achieves the optimization goal while adhering to the stated constraints.
\section{Evaluation}
\label{sec:eval}

\subsection{Experimental Setup}


We evaluated various DSE methods on both the \textit{roofline model} and the \textit{LLMCompass model}~\cite{llmcompass2024isca}. 
LLMCompass is a GPU simulator specialized for LLM inference workloads; it models software-hardware co-optimizations and achieves performance and area estimations within 10\% relative error compared to real hardware. 
We extended LLMCompass to include critical path analysis, enabling identification of dominant stalls for both TTFT and TPOT metrics, with negligible overhead.


\subsection{DSE Benchmark}
We evaluate several state-of-the-art open-source LLMs, including Qwen3-Next-80B-A3B-Instruct (Qwen-3)~\cite{qwen3}, Phi-4-reasoning (Phi-4)~\cite{phi4} and Llama-3.1-8B-Instruct (Llama-3.1)~\cite{llama3_8b_instruct}.
For brevity, we refer to the models using the shortened names shown in parentheses throughout the paper.

The benchmark comprises 308 questions for bottleneck analysis, 127 for performance/area prediction, and 30 for parameter tuning.
We measure LLMs' DSE capability using the accuracy metrics, which is computed as the proportion of correctly solved problems across the dataset.

As shown in Table~\ref{tab:passk}, Qwen-3 achieved the highest accuracy among the evaluated models under the default system prompt, which already provides the necessary architectural context. However, the substantial variability across tasks indicates that additional error-handling mechanisms are required.
To understand these gaps, we analyzed the model’s systematic failure patterns and translated them into explicit corrective rules. With this additional human-crafted guidance, Qwen-3 achieved noticeably higher accuracy, as reported in the \textit{Accuracy (Enhanced)} column.
In \lumina, these rules are integrated into the \textit{Strategy Engine}, enabling the system to enforce consistent and reliable architectural reasoning during design-space exploration.

\begin{table}[htbp]
\centering
\caption{Accuracy across tasks and open-source LLMs}
\footnotesize
\label{tab:passk}
\setlength{\tabcolsep}{3pt}
\renewcommand{\arraystretch}{1}

\begin{tabular}{l l cc}
\toprule
Benchmark Task & Model & Accuracy (Original) & Accuracy (Enhanced) \\
\midrule

\multirow{3}{*}{Bottleneck Analysis}
  & Phi-4       & 0.70 & 0.76 \\
  & Qwen-3      & \textbf{0.73} & \textbf{0.80} \\
  & Llama-3.1  & 0.47 & 0.53 \\
\midrule

\multirow{3}{*}{Perf/Area Prediction}
  & Phi-4       & 0.42 & 0.61 \\
  & Qwen-3      & \textbf{0.59} & \textbf{0.82} \\
  & Llama-3.1  & 0.23 & 0.39 \\
\midrule

\multirow{3}{*}{Parameter Tuning}
  & Phi-4       & 0.30 & 0.48  \\
  & Qwen-3      & \textbf{0.40} & \textbf{0.63} \\
  & Llama-3.1  & 0.26 & 0.46 \\
\bottomrule
\end{tabular}
\end{table}

For \textit{bottleneck analysis}, models often selected multi-resource configurations containing irrelevant parameters instead of targeting the resource most correlated with the stall. We therefore constrained the \textit{Strategy Engine} to \textbf{focus solely on the dominant bottleneck}. A further limitation is that it fails to recognize the adverse effects of enlarging resources such as the systolic array height and width, which may cause significant compute under-utilization. Mitigating this gap requires enabling the model to reason from prior experience or fine-tuning it on datasets that capture such design pitfalls. 

For \textit{performance and area prediction}, models frequently computed deltas against a zero baseline rather than the defined sensitivity reference. To ensure consistency, the \textit{Strategy Engine} was required to \textbf{always compute deltas relative to the sensitivity reference}.

For \textit{parameter tuning}, models often attempt to compensate for an unresolved dominant bottleneck by adjusting multiple non-critical resources, increasing reasoning complexity beyond the model's capability.  
To address this, we instructed the \textit{Strategy Engine} to \textbf{adjust only the least critical resource} when mitigating the dominant stall.

\subsection{\lumina Framework}

We evaluate the DSE methods under a GPT-3 inference workload using the full 4.7-million design space.
We use the NVIDIA A100 as the reference design (Table~\ref{table:lumina_designs}) and adopt 8-way tensor parallelism as the deployment strategy.
TTFT latency is measured by executing a single GPT-3 layer with a batch size of 8 and an input sequence length of 2048.
TPOT latency is defined as the time to generate the 1024th output token under the same configuration.
All operators are executed in FP16 precision.
This setup ensures that both TTFT and TPOT characteristics are faithfully captured, providing a realistic basis for DSE evaluation.
The comparison metrics are Pareto Hyper Volume (PHV)\cite{archexplorer2023micro} and Sample Efficiency, defined as number of points that are better than the reference point in all objectives divides the total exploration samples.
Sample efficiency measures the fraction of evaluated designs that improve over the reference point; higher values indicate a more efficient exploration of the design space.

\begin{figure}[htbp]
    \centering
    \includegraphics[width=1\linewidth]{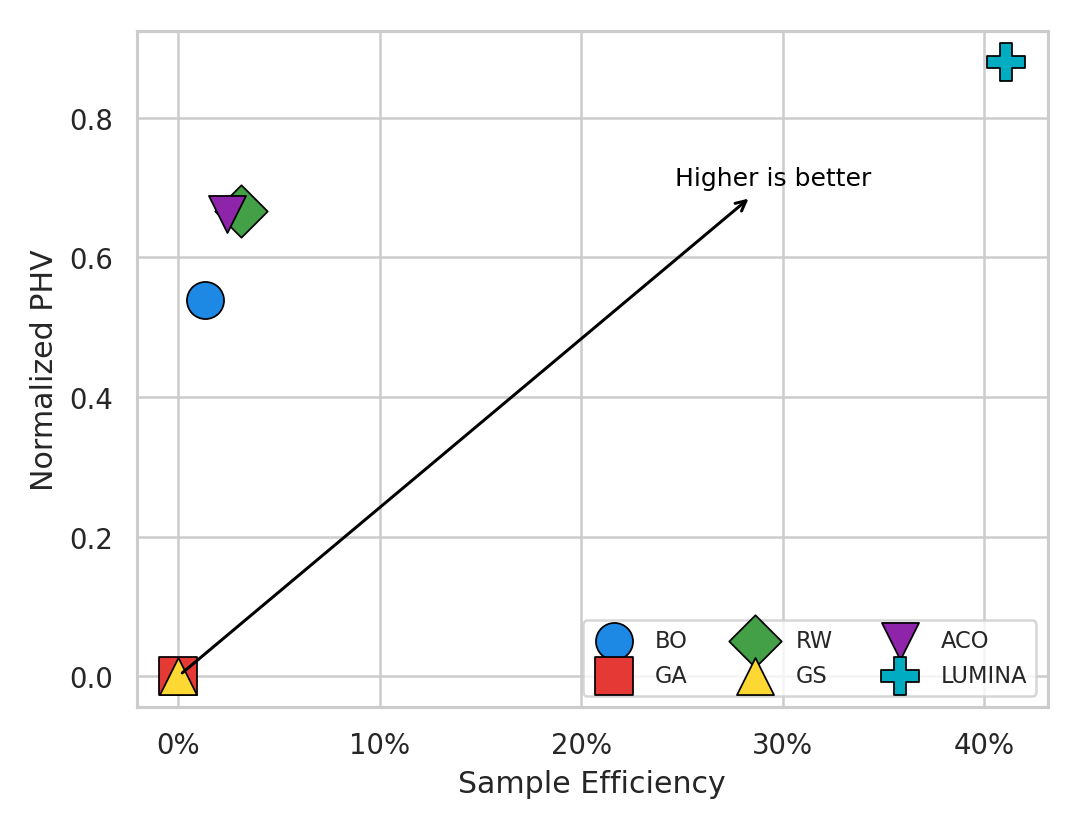}
    \caption{Mean PHV vs. Sample Efficiency among DSE Methods.}
    \label{fig:hv_se_mean}
\end{figure}

\begin{figure}[t]
    \centering
    \includegraphics[width=1\linewidth]{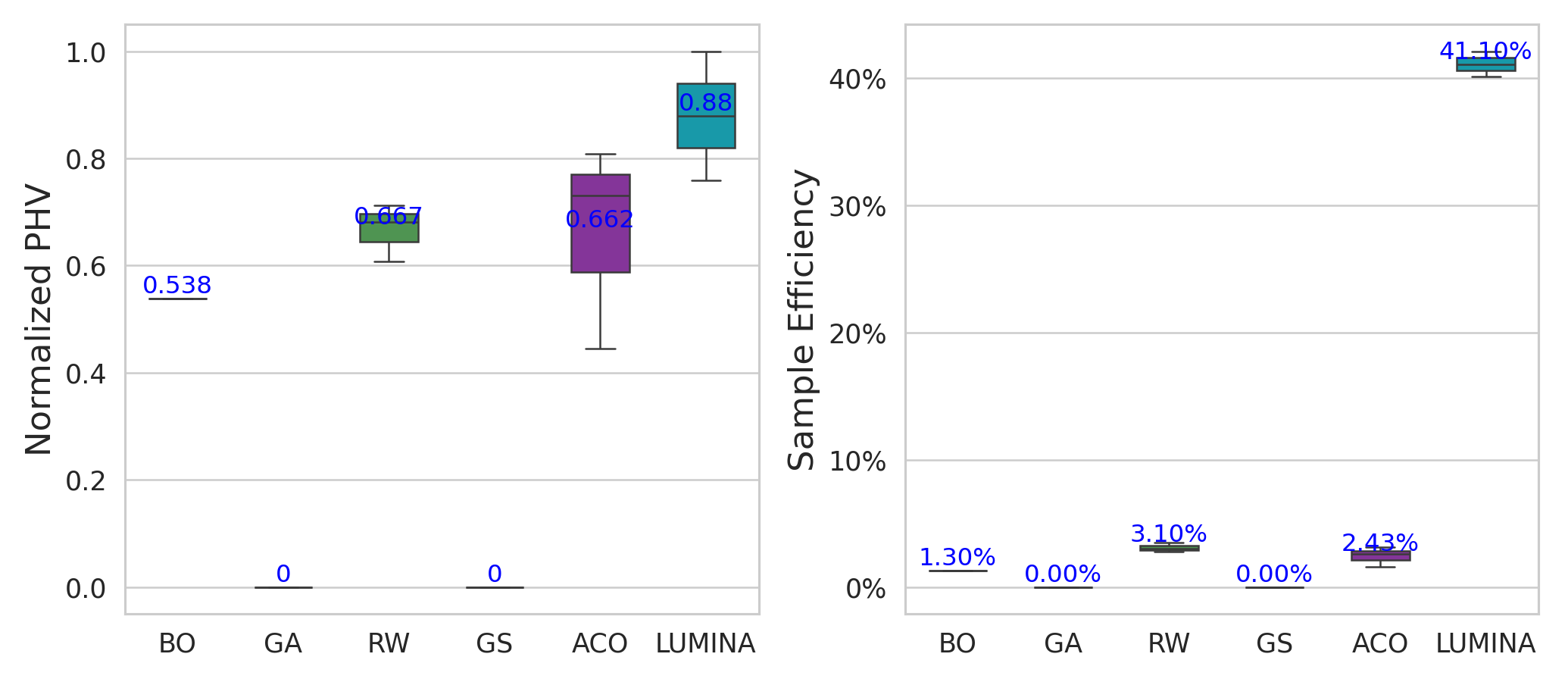}
    \caption{Distribution of PHV vs. Sample Efficiency among DSE Methods.}
    \label{fig:hv_se_dist}
\end{figure}

We begin by comparing the PHV, associated variance, and sample efficiency of each method across 1,000 samples and multiple independent trials under roofline model evaluation. 
As shown in Figure~\ref{fig:hv_se_mean}, \textbf{\lumina achieves the highest mean PHV and sample efficiency}, outperforming all other methods by \textbf{32.9\% in PHV} and \textbf{17.5$\times$ in sample efficiency}. Among the black-box baselines, \textbf{ACO} and \textbf{RW} exhibit comparable PHV and sample efficiency, primarily due to their \textit{chance sampling} behavior, as illustrated in Figure~\ref{fig:hv_se_dist}. Notably, \textbf{ACO’s best-to-worst normalized PHV ratio reaches up to 1.82x}, indicating its high variability across runs. \textbf{BO} delivers consistently strong performance, ranking fourth overall, whereas \textbf{GA} and \textbf{GS} consistently fail to discover high-quality designs. Furthermore, \textbf{\lumina maintains superior sample efficiency across multiple runs}, highlighting its stability and robustness.

\begin{figure}[htbp]
    \centering
    \includegraphics[width=1\linewidth]{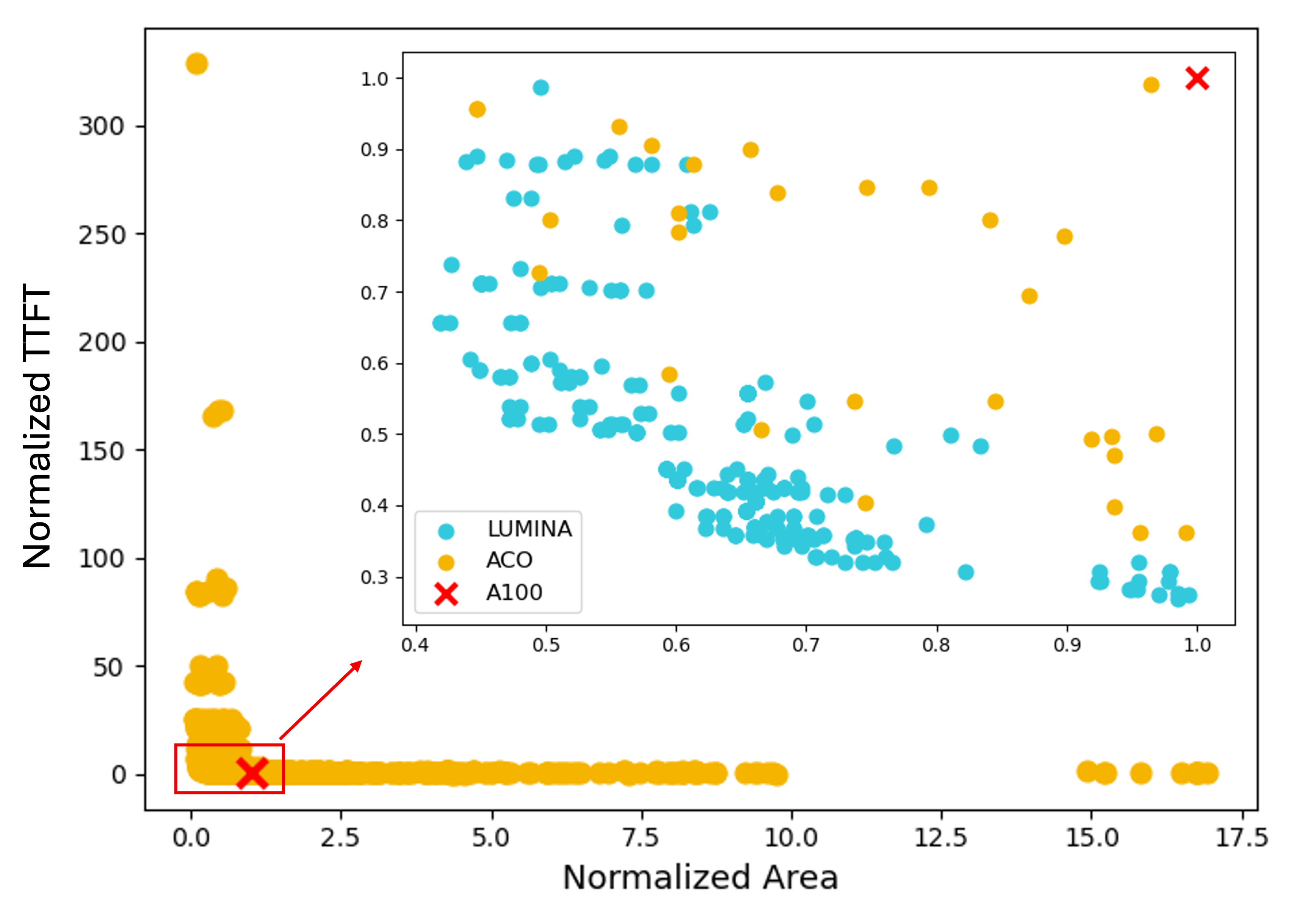}
    \caption{Search Pattern Comparison between ACO and \lumina}
    \label{fig:traj_bo_lumina}
\end{figure}

Next, we analyze the \textbf{DSE pattern} to show \textit{why} \lumina outperforms other methods. 
As shown in Figure~\ref{fig:traj_bo_lumina}, \textbf{\lumina} explores the design space far more efficiently by leveraging its understanding of the simulator’s internal mechanisms. 
In contrast, the state-of-the-art black-box baseline, \textbf{ACO}, adopts a far-to-near exploration strategy, consuming numerous samples merely to map the objective space before reaching promising regions. 
As a result, \lumina identifies substantially more high-quality design points within the same sample budget—specifically, 421 superior designs within 1,000 samples, compared to only 24 for ACO—\textbf{demonstrating its significant advantage in sample efficiency}.

Then, we evaluated all methods on the \textbf{LLMCompass model}. Owing to the high simulation cost of end-to-end inference evaluation, we imposed a strict sample budget of only 20 evaluations, which requires approximately one week to complete. Under this constraint, none of the black-box algorithms succeeded in identifying any design that surpassed the reference point of NVIDIA A100~\cite{llmcompass2024isca}. This failure primarily stems from the insufficient number of samples, which prevents these methods from effectively mapping the high-dimensional design space. \textbf{In contrast, \lumina was the \textit{only} methodology that successfully discovered \textit{six} designs surpassing the reference point}, demonstrating its unique effectiveness under extreme sample constraints.

\begin{table}[htbp]
\centering
\footnotesize
\caption{Top-2 designs identified by \lumina Compared with NVIDIA A100}
\label{table:lumina_designs}
\renewcommand{\arraystretch}{1}
\setlength{\tabcolsep}{5pt}
\begin{tabular}{|l|c|c|c|}
\hline
\textbf{Specifications} & \textbf{Design A} & \textbf{Design B} & \textbf{A100} \\
\hline
Interconnect Link Count & \textbf{24} & \textbf{18} & 12\\
Core Count & \textbf{64} & \textbf{96} & 108\\
Sublane Count & 4 & 4 & 4\\
Vector Width & \textbf{16} & \textbf{16} & 32\\
Systolic Array Height$\times$Weight & \textbf{32$\times$32} & \textbf{32$\times$32} & 16$\times$16\\
SRAM Size (KB) & 128 & 128 & 128\\
Global Buffer (MB) & 40 & 40 & 40\\
Memory Channel Count & \textbf{6} & \textbf{6} & 5\\
\hline
Normalized TTFT & 0.717 & \textbf{0.592} & 1.000\\
Normalized TPOT & \textbf{0.947} & 0.948 & 1.000\\
Normalized Area & 0.772 & 0.952 & 1.000\\
\hline
TTFT/Area & \textbf{1.805} & 1.366 & 1.000\\
TPOT/Area & \textbf{1.770} & 1.107 & 1.000\\
\hline
\end{tabular}
\end{table}

Finally, we analyze the rationale behind the top-performing designs identified by \textbf{\lumina} relative to the NVIDIA A100 baseline, as shown in Table~\ref{table:lumina_designs}. \lumina consistently reallocates architectural resources to address the dominant performance limits, demonstrating that co-optimizing compute hierarchy, memory bandwidth, and interconnect bandwidth yields substantial gains.

Concretely, it increases the \textbf{interconnect link count} (12 → 18/24) to boost communication capability, while offsetting the area cost through a moderate reduction in \textbf{core count} (108 → 64/96). Both optimized designs retain a wide \textbf{systolic array} ($32\times32$) and add one additional \textbf{memory channel} (5 → 6), thereby improving memory bandwidth and sustaining compute throughput.

\textbf{Design~A} delivers the best overall trade-off, achieving \textbf{1.805$\times$} TTFT/Area efficiency and \textbf{1.77$\times$} TPOT/Area efficiency relative to A100, while using only \textbf{77\%} of its area. \textbf{Design~B} attains comparable TPOT and reduces normalized TTFT to \textbf{0.592}, making it suitable for strict TTFT-oriented SLAs. These outcomes show that \lumina effectively identifies balanced design points where compute and communication resources are jointly optimized to maximize utilization.

\section{Conclusion}
\label{sec:conclusion}

In this paper, we introduced \textbf{\lumina}, a novel LLM-Guided GPU architecture exploration framework; a \textbf{DSE benchmark} that systematically evalutes LLM's capability on DSE tasks and ensures consistent architectural reasoning; and two \lumina-discovered designs and their insights.


\newpage
\bibliographystyle{ACM-Reference-Format}
\bibliography{refs}

\end{document}